\documentclass[aps,prl,preprintnumbers,amsmath,amssymb,twocolumn,superscriptaddress,showpacs,floatfix]{revtex4}

\usepackage{graphicx}% Include figure files
\usepackage{dcolumn}% Align table columns on decimal point
\usepackage{bm}% bold math
\usepackage{color}
\usepackage{epsfig}
\begin{document}

\title{Twist-4 contributions to the azimuthal asymmetry in SIDIS}
\author{Yu-kun Song}
\affiliation{School of Physics, Shandong University, Jinan, Shandong 250100, China}
\author{Jian-hua Gao}
\affiliation{Department of Modern Physics, University of Science and Technology of China, Hefei, Anhui 230026, China}
\author{Zuo-tang Liang}
\affiliation{School of Physics, Shandong University, Jinan, Shandong 250100, China}
\author{Xin-Nian Wang}
\affiliation{Nuclear Science Division, MS 70R0319, Lawrence Berkeley
National Laboratory, Berkeley, California 94720}

\date{\today}

%\preprint{LBNL-}

%\vspace{-1.5in}
%\vspace{1.4in}

\begin{abstract}
We calculate the differential cross section for the unpolarized semi-inclusive deeply inelastic scattering (SIDIS)
process $e^-+N \to e^-+q+X$ in leading order (LO) of perturbative QCD and up to twist-4 in power corrections
and study in particular the azimuthal asymmetry $\langle\cos2\phi\rangle$. The final results are expressed in 
terms of transverse momentum dependent (TMD) parton matrix elements of the target nucleon 
up to twist-4. 
%Under the maximal two-gluon correlation approximation, these TMD parton matrix elements  in a nucleus
%can be expressed terms of a Gaussian convolution of that in a nucleon with the width given by the jet transport
%parameter inside cold nuclei. 
We also apply it to $e^-+A \to e^-+q+X$ and illustrate numerically the nuclear dependence of the
azimuthal asymmetry $\langle\cos2\phi\rangle$ by using a Gaussian ansatz for the
TMD parton matrix elements.
 
\end{abstract}


\maketitle

%\begin{multicols}{2}

\section{Introduction}

Inclusive and semi-inclusive deep inelastic scatterings (SIDIS) are important tools to understand
the structure of nucleon and nucleus governed by the
Quantum Chromodynamics (QCD) for the strong interaction.
The azimuthal asymmetries and their spin and/or nuclear dependences
of the SIDIS cross sections are directly related to the parton distribution and polarization inside nucleon
or nuclei and therefore are the subjects of intense studies both 
theoretically\cite{Georgi:1977tv,Cahn:1978se,Berger:1979kz,Liang:1993re,
Mulders:1995dh,Oganesian:1997jq,Chay:1997qy,Nadolsky:2000kz,
Anselmino:2006rv,Anselmino:2005nn,Liang:2006wp,Gao:2010mj}
and experimentally\cite{Aubert:1983cz,Arneodo:1986cf,Adams:1993hs,
Breitweg:2000qh,Chekanov:2002sz,Airapetian:2001eg,Airapetian:2002mf,Airapetian:2004tw,
Alexakhin:2005iw,Webb:2005cd,:2008rv,Alekseev:2010dm}.
They provide us with a glimpse into the dynamics of strong interaction within nucleons or nuclei 
and a baseline for the study of parton dynamics in other extreme conditions at high temperature and
baryon density.

In the unpolarized SIDIS experiments, the azimuthal angle $\phi$  of the final hadrons is defined with respect to 
the leptonic plane and is directly related to the transverse momentum of the hadron from either parton 
fragmentation or the initial and final state interaction of the parton before hadronization.
In this paper we will restrict our study to SIDIS $e^-+N (A) \to e^-+q+X$ of quark jet production so that we
don't need to deal with the azimuthal asymmetry resulting from parton fragmentation 
and have no need to consider Boer-Mulders effect \cite{Boer:1997nt}. 
We instead focus primarily on the effect of initial and final state interaction.
In the large transverse momentum region, the azimuthal asymmetries arise predominately
from hard gluon bremsstrahlung that can be calculated using perturbative QCD (pQCD) \cite{Georgi:1977tv},
and are clearly observed in experiments\cite{Aubert:1983cz,Arneodo:1986cf,Adams:1993hs,Breitweg:2000qh,Chekanov:2002sz}.
On the other hand, in the small transverse momentum region  $p_{h\perp}\sim k_\perp \le 1$GeV/c,
the asymmetry was shown\cite{Cahn:1978se} to arise mainly from the intrinsic
transverse momentum of quarks in nucleon and is
a higher twist effect proportional to $k_\perp/Q$ for $\langle\cos\phi\rangle$
and to $k_\perp^2/Q^2$ for $\langle\cos 2\phi\rangle$.
(Here, $p_{h\perp}$ denotes the transverse momentum of the hadron produced,
$k_\perp$ is the intrinsic transverse momentum of quark in nucleon,
$Q^2=-q^2$ and $q$ is the four-momentum transfer from the lepton).
The calculations in \cite{Cahn:1978se} are based on a generalization
of the naive parton model to include intrinsic transverse momentum.
To go beyond the naive parton model, one has to consider multiple soft
gluon interaction between the struck quark and the remanent of the target nucleon or nucleus.
Inclusion of such soft gluon interaction ensures the gauge invariance of the final
results and relate the azimuthal asymmetry to the transverse momentum
dependent (TMD) parton matrix elements of the nucleon or nucleus.

Within the framework of TMD parton distributions and correlations, the intrinsic
transverse momentum of partons arises naturally from multiple soft
gluon interaction inside the nucleon or nucleus. The TMD parton distributions and correlations
can be in fact expressed in terms of the expectation values of matrix elements
related to the accumulated total transverse momentum as a result of the
color Lorentz force enforced upon the parton through soft gluon exchange \cite{Liang:2008vz}.
These soft gluon interactions are responsible for the single-spin asymmetries
observed in SIDIS,  $pp$ and $\bar pp$ collisions. They also lead to the 
transverse momentum broadening \cite{Liang:2008vz} of hadron production in 
deep-inelastic lepton-nucleus scattering\cite{VanHaarlem:2007kj,VanHaarlem:2009zz,Domdey:2008aq}
as well as the jet quenching observed at the Relativistic Heavy Ion Collider (RHIC) \cite{Gyulassy:1993hr,Baier:1996sk,
Wiedemann:2000za,Gyulassy:2000er,Guo:2000nz,Wang:2001ifa}.  Such transverse
momentum broadening inside nucleus is directly related to the gluon 
saturation scale \cite{Liang:2008vz,McLerran:1993ka} and can be studied directly through the nuclear 
dependence of the azimuthal asymmetry in SIDIS.

Higher twist contributions in inclusive DIS have been studied systematically using the
collinear expansion technique \cite{Ellis:1982wd,Qiu:1988dn,Qiu:1990xxa} which not only provides 
a useful tool to study the higher twist contributions but also is a necessary procedure to ensure
gauge invariance of the parton distribution and/or correlation functions.
In Ref.\cite{Liang:2006wp}, such collinear expansion is extended to the SIDIS process $e^-+N\to e^-+q+X$
and calculation of the TMD differential cross section and the azimuthal asymmetries up to twist-3.
Taking into account of multiple gluon scattering, the study found the azimuthal  asymmetry $\langle \cos\phi\rangle$
proportional to a twist-3 TMD parton correlation 
function $f_{q\perp}(x,k_\perp)$ defined as,
%\begin{widetext}
\begin{eqnarray}
f_{q\perp}^N(x,k_\perp) &&
=\int \frac{p^+dy^- d^2y_{\perp}}{(2\pi)^3}
e^{ix p^+ y^- -i\vec{k}_{\perp}\cdot \vec{y}_{\perp}} \nonumber \\
&& \times\langle N|\bar{\psi}(0)\frac{\slash{\hspace{-5pt}k_\perp}}{2k_\perp^2}{\cal{L}}(0;y)\psi(y)|N\rangle,
\label{eq:fqperp}
\end{eqnarray}
where ${\cal{L}}(0;y)$ is the gauge link,
\begin{eqnarray}
\label{CollinearGL}
&&{\cal{L}}(0;y)=\mathcal{L}^\dag_{\parallel}(\infty,\vec{0}_\perp; 0,\vec{0})
\mathcal{L}_\perp(\infty,\vec 0_\perp;\infty,\vec{y}_{\perp}) \nonumber\\
&&\phantom{{\cal{L}}(0;y)=}
\mathcal{L}_{\parallel}(\infty,\vec{y}_{\perp}; y^-,\vec{y}_{\perp},\vec{y}_\perp),\nonumber\\
\label{TMDGL}
&&{\cal{L}}_{\parallel}(\infty,\vec{y}_{\perp}; y^-,\vec y_\perp)\equiv 
%P \exp \left[- i g \int_{y^-}^\infty d \xi^{-} A^+ ( \xi^-, \vec{y}_{\perp})\right] , \nonumber \\
P e^{- i g \int_{y^-}^\infty d \xi^{-} A^+ ( \xi^-, \vec{y}_{\perp})} , \nonumber \\
&&\mathcal{L}_\perp(\infty,\vec 0_\perp;\infty,\vec{y}_{\perp}) \equiv  
Pe^{-ig\int_{\vec 0_\perp}^{\vec y_\perp} d\vec\xi_\perp\cdot
\vec A_\perp(\infty,\vec\xi_\perp)},
\end{eqnarray}
%\end{widetext}
from the resummation of multiple soft gluon interaction that ensures the gauge invariance of the
twist-3 parton correlation function in Eq.~(\ref{eq:fqperp}) under any gauge transformation.
The asymmetry obtained within this generalized collinear expansion method  reduces to that in the
naive parton model \cite{Cahn:1978se} if and only if one neglects the soft gluon interaction as contained
in the gauge link or equivalently by setting the strong coupling constant $g=0$ in the final result.
Measurements of $\langle \cos\phi\rangle$ in $e^-+N\to e^-+q+X$ and its $k_\perp$-dependence
therefore provide an unique determination of this new parton correlation function in Eq.~(\ref{eq:fqperp}).
Furthermore, the nuclear dependence of the asymmetry \cite{Liang:2008vz} from multiple soft gluon
interaction within the target nucleus can probe the transverse momentum broadening or the jet
quenching parameter in cold nuclear matter \cite{Gao:2010mj} which also determines the gluon
saturation scale in cold nuclei.

In this paper, we present a complete calculation of the hadronic tensor and the differential cross section
for $e^-+N\to e^-+q+X$ up to twist-4. We study in particular the azimuthal asymmetry  $\langle\cos2\phi\rangle$
in terms of the corresponding TMD quark correlation functions and its nuclear dependence.
For completeness, in Sec. II, we present the formulae for calculating the hadronic tensor and
differential cross sections within the framework of generalized collinear expansion.
In Sec. III,  we present the cross section and discuss azimuthal asymmetry  $\langle\cos2\phi\rangle$
including its nuclear dependence with a Gaussian ansatz for the TMD correlation functions. 
A summary is given in Sec. IV.

\section{Hadronic tensor $W_{\mu\nu}$ in $e^-+N\to e^-+q+X$ up to twist-4}

We consider the SIDIS process $e^-+N\to e^-+q+X$ with unpolarized beam and target.
The differential cross section is given by,
%\textcolor{blue}{
\begin{equation}
d\sigma=\frac{\alpha_{em}^2e_q^2}{sQ^4}L^{\mu\nu}(l,l')\frac{d^2W_{\mu\nu}}{d^2k'_\perp}
\frac{d^3l'd^2k'_\perp}{ 2E_{l'}},
\end{equation}
where $l$ and $l'$ are respectively the four-momenta of the incoming and outgoing leptons,
$p$ is the four-momentum of the incoming nucleon $N$,
%$q$ is the four momentum transfer,
$k'$ is the four-momentum of the outgoing quark.
We neglect the masses and use the light-cone coordinates.
The unit vectors are taken as,
$\bar n^\mu=(1,0,0,0)$, $n^\mu=(0,1,0,0)$, $n_{\perp 1}^\mu=(0,0,1,0)$,
$n_{\perp 2}^\mu=(0,0,0,1)$.
We chose the coordinate system in the way so that,
$p=p^+\bar n$, $q=-x_Bp+nQ^2/(2x_Bp^+)$, $l_\perp=|\vec l_\perp|n_{\perp 1}$,
and $k_\perp=(0,0,\vec k_\perp)$;
where $x_B=Q^2/2p\cdot q$ is the Bjorken-$x$ and $y=p\cdot q/p\cdot l$.
The leptonic tensor $L^{\mu\nu}$ is defined as usual ,
\begin{equation}
L^{\mu\nu}(l,l')=4[l^\mu{l'}^\nu+l^\nu{l'}^\mu-(l\cdot l')g^{\mu\nu}],
\end{equation}
and the differential hadronic tensor is,
%\textcolor{blue}{
\begin{equation}
\frac{d^2W_{\mu\nu}}{d^2k'_\perp}=\int \frac{dk'_z}{(2\pi)^3 2E_{k'}}W_{\mu\nu}^{(si)}(q,p,k'),
\end{equation}
%where $W_{\mu\nu}^{(si)}$ for SIDIS is defined as,
\begin{eqnarray}
W_{\mu\nu}^{(si)}(q,p,k')&=&
\frac{1}{2\pi}\sum_X \langle N| J_\mu(0)|k',X\rangle \langle k',X| J_\nu(0)|N\rangle \nonumber\\
&&\times (2\pi)^4\delta^4(p+q-k'-p_X),
\end{eqnarray}
where the superscript $(si)$ denotes SIDIS.
It has been shown\cite{Liang:2006wp} that, after  collinear expansion,
the hadronic tensor can be expressed in an expansion series characterized by the number of
covariant derivatives in the parton matrix elements in each term,
\begin{equation}
\frac{d^2W_{\mu \nu}}{d^2{k}_\perp}=\sum_{j=0}^\infty
\frac{d^2\tilde W^{(j)}_{\mu \nu}}{d^2{k}_\perp},
\end{equation}
%where the number $j$ in the superscript of $\tilde W_{\mu\nu}$ denotes the number of gluon(s)
%involved in the multiple gluon scattering and the tilde above $W$
%denotes the results after collinear expansion.
%They are given by,
\begin{widetext}
\begin{eqnarray}
\label{eq:Wsi0}
&&\frac{d\tilde W_{\mu\nu}^{(0)}}{d^2k'_\perp}
=\frac{1}{2\pi} \int dx d^2k_\perp
{\rm Tr}[\hat H_{\mu\nu}^{(0)}(x)\ \hat \Phi^{(0)N}(x,k_\perp)] \delta^{(2)}(\vec k_\perp-\vec{k'}_\perp);\\
%%%%%%%
\label{eq:Wsi1}
&&\frac{d\tilde W_{\mu\nu}^{(1)}}{d^2k'_\perp}
=\frac{1}{2\pi}
\int dx_1d^2k_{1\perp} dx_2d^2k_{2\perp} \sum_{c=L,R}
{\rm Tr}\bigl[\hat H_{\mu\nu}^{(1,c)\rho}(x_1,x_2) \omega_\rho^{\ \rho'}
\hat \Phi^{(1)N}_{\rho'}(x_1,k_{1\perp},x_2,k_{2\perp})\bigr] \delta^{(2)}(\vec k_{c\perp}-\vec{k'}_\perp); \phantom{XX}\\
%%%%%%%%
\label{eq:Wsi2}
&&\frac{d\tilde W_{\mu\nu}^{(2)}}{d^2k'_\perp}=\frac{1}{2\pi}
\int dx_1d^2k_{1\perp}dx_2d^2k_{2\perp}dxd^2k_{\perp}\nonumber\\
&&\phantom{\frac{d\tilde W_{\mu\nu}^{(2)}}{d^2k'_\perp}=\frac{1}{2\pi}}
\sum_{c=L,R,M}{\rm Tr}\bigl[\hat H_{\mu\nu}^{(2,c)\rho\sigma}(x_1,x_2,x)
\omega_\rho^{\ \rho'}\omega_\sigma^{\ \sigma'} \hat\Phi^{(2)N}_{\rho'\sigma'}(x_1,k_{1\perp},x_2,k_{2\perp},x,k_\perp)\bigr]
\delta^{(2)}(\vec k_{c\perp}-\vec{k'}_\perp).\
\end{eqnarray}
where, for different cuts $c=L$, $R$ or $M$, $\vec k_{c\perp}$ denotes
$\vec k_{L\perp}=\vec k_{1\perp}$, $\vec k_{R\perp}=\vec k_{2\perp}$,  and
$\vec k_{M\perp}=\vec k_{\perp}$;
$\omega_\rho^{\ \rho'}=g_\rho^{\ \rho'}-\bar n_\rho n^{\rho'}$ is a projection operator.
The matrix elements are defined as,
\begin{eqnarray}
\hat\Phi^{(0)N}(x,k_\perp)&=&\int \frac{p^+dy^- d^2y_\perp}{(2\pi)^3}
e^{ix p^+ y^- -i\vec{k}_{\perp}\cdot \vec{y}_{\perp}}
\langle N|\bar{\psi}(0){\cal{L}}(0;y)\psi(y)|N\rangle,% \phantom{XXXXXXXXXXXXXXXXXXXX}
\label{eq:Phi0def}\\
%%%%%%
\hat\Phi^{(1)N}_\rho(x_1,k_{1\perp},x_2,k_{2\perp})
&=&\int \frac{p^+dy^- d^2y_\perp}{(2\pi)^3}\frac{p^+dz^-d^2z_\perp}{(2\pi)^3}
e^{ix_2p^+z^- -i\vec{k}_{2\perp}\cdot \vec{z}_{\perp}+
ix_1p^+(y^--z^-)-i\vec{k}_{1\perp}\cdot (\vec y_\perp-\vec z_\perp)} \nonumber\\
&&\langle N|\bar\psi(0) {\cal L}(0;z)D_\rho(z){\cal L}(z;y)\psi(y)|N\rangle,
\label{eq:Phi1def}\\
%%%%%%
\hat\Phi^{(2)N}_{\rho\sigma}(x_1,k_{1\perp},x_2,k_{2\perp},x,k_\perp)&=&
\int \frac{p^+dz^-d^2z_\perp}{(2\pi)^3} \frac{p^+dy^- d^2y_\perp}{(2\pi)^3} \frac{p^+d{y'}^- d^2{y'}_\perp}{(2\pi)^3}\nonumber\\
&&e^{ix_2p^+z^- -i\vec{k}_{2\perp}\cdot \vec{z}_{\perp}+
ixp^+({z'}^--z^-)-i\vec{k}_{\perp}\cdot (\vec {z'}_\perp-\vec {z}_\perp)+
ix_1p^+({y}^--{z'}^-)-i\vec{k}_{1\perp}\cdot (\vec {y}_\perp-\vec {z'}_\perp)} \nonumber\\
&&\langle N|\bar\psi(0){\cal L}(0;z) D_\rho(z) {\cal L}(z;z')D_\sigma(z'){\cal L}(z';y)\psi(y)|N\rangle,
\label{eq:Phi2def}
\end{eqnarray}
where ${\cal{L}}(0;y)$ is the gauge link as defined in Eq.~(\ref{eq:fqperp}), and also in the remainder of this paper, for brevity, 
unless explicitly specified, the coordinate $y$ in the field operator denotes $(0,y^-,\vec{y}_{\perp})$.

The hard parts after the collinear expansion are given as \cite{Liang:2006wp},
\begin{eqnarray}
\hat H_{\mu\nu}^{(0)}(x)&=&\frac{2\pi}{2q\cdot p}
\gamma_\mu(\slash{\hspace{-5pt}q}+x\slash{\hspace{-5pt}p})\gamma_\nu\delta(x-x_B),\\
\hat H_{\mu\nu}^{(1,L)\rho}(x_1,x_2)&=&\frac{2\pi}{(2q\cdot p)^2}
\frac{\gamma_\mu(\slash{\hspace{-5pt}q}+x_2\slash{\hspace{-5pt}p})\gamma^\rho(\slash{\hspace{-5pt}q}+x_1\slash{\hspace{-5pt}p})
\gamma_\nu}{x_2-x_B-i\varepsilon}\delta(x_1-x_B),\\
%%%%%%%%%%%%%%%
\hat H_{\mu\nu}^{(2,L)\rho\sigma}(x_1,x_2,x)&=&\frac{2\pi}{(2q\cdot p)^3}
\frac{\gamma_\mu(\slash{\hspace{-5pt}q}+x_2\slash{\hspace{-5pt}p})\gamma^\rho(\slash{\hspace{-5pt}q}+x\slash{\hspace{-5pt}p})
\gamma^\sigma(\slash{\hspace{-5pt}q}+x_1\slash{\hspace{-5pt}p})\gamma_\nu}
{(x-x_B-i\varepsilon)(x_2-x_B-i\varepsilon)}\delta(x_1-x_B).
\end{eqnarray}
These equations form the basis for calculating the hadronic tensor in $e^-+N\to e^-+q+X$.
Due to the existence of the projection operators $\omega_\rho^{\ \rho'}$ and $\omega_\sigma^{\ \sigma'}$,
the hard parts can be simplified to,
\begin{eqnarray}
&&\hat H^{(0)}_{\mu\nu}(x)=\pi\hat h^{(0)}_{\mu\nu}\delta(x-x_B),
\label{eq:H0simple}\\
&&\hat H^{(1,L)\rho}_{\mu\nu}(x_1,x_2)\omega_\rho^{\ \rho'}
=\frac{\pi}{2q\cdot p}\hat h^{(1)\rho}_{\mu\nu}\omega_\rho^{\ \rho'}\delta(x_1-x_B),
\label{eq:H1Lsimple}\\
%&&\hat H^{(1,R)\rho}_{\mu\nu}(x_1,x_2)\omega_\rho^{\ \rho'}=
%\frac{\pi}{2q\cdot p}\gamma_0\hat h^{(1)\rho\dag}_{\nu\mu}\gamma_0\omega_\rho^{\ \rho'}\delta(x_2-x_B),
%\label{eq:H1Rsimple} \\
%%%%%%%%%%%%%%%%%%%%%%
&&\hat H^{(2,L)\rho\sigma}_{\mu\nu}(x_1,x_2,x)\omega_\rho^{\ \rho'}\omega_\sigma^{\ \sigma'}=
\frac{2\pi}{(2q\cdot p)^2}\bigl[\bar n^\rho\hat h^{(1)\sigma}_{\mu\nu}+\frac{\hat N^{(2)\rho\sigma}_{\mu\nu}}{x_2-x_B-i\varepsilon}\bigr]
\omega_\rho^{\ \rho'}\omega_\sigma^{\ \sigma'}\delta(x_1-x_B),
\label{eq:H2Lsimple}\\
%%%%%%%%%%%%%%%%%%
%&& \hat H^{(2,R)\rho\sigma}_{\mu\nu}(x_1,x_2,x)\omega_\rho^{\ \rho'}\omega_\sigma^{\ \sigma'}=
%\frac{2\pi}{(2q\cdot p)^2}\bigl[p^\rho\gamma_0\hat h^{(1)\sigma\dag}_{\nu\mu}\gamma_0+
%\frac{q^-\hat N^{(2)\rho\sigma}_{\mu\nu}}{x_1-x_B+i\varepsilon}\bigr]
%\omega_\rho^{\ \rho'}\omega_\sigma^{\ \sigma'}\delta(x_2-x_B),
%\label{eq:H2Rsimple}\\
%%%%%%%%%%%%%%%%%%%%
&& \hat H^{(2,M)\rho\sigma}_{\mu\nu}(x_1,x_2,x)\omega_\rho^{\ \rho'}\omega_\sigma^{\ \sigma'}=
\frac{2\pi }{(2q\cdot p)^2}\hat h^{(2)\rho\sigma}_{\mu\nu}
%\frac{2\pi p^+}{(2q\cdot p)^2}\bigl[\bar n^\rho\hat h^{(1)\sigma}_{\mu\nu}-\bar n^\sigma\hat h^{(1)\rho}_{\mu\nu}
%-2\bar n^\sigma\gamma_\mu\gamma^\rho\gamma_\nu+\hat{\bar N}^{(2)\sigma\rho}_{\mu\nu}\bigr]
\omega_\rho^{\ \rho'}\omega_\sigma^{\ \sigma'}\delta(x-x_B),
\label{eq:H2Msimple}
\end{eqnarray}
where $\hat h^{(0)}_{\mu\nu}=\gamma_\mu\slash{\hspace{-5pt}n}\gamma_\nu/p^+$,
$\hat h^{(1)\rho}_{\mu\nu}=\gamma_\mu\slash{\hspace{-5pt}\bar n}\gamma^\rho\slash{\hspace{-5pt}n}\gamma_\nu$,
$\hat h^{(2)\rho\sigma}_{\mu\nu}=p^+\gamma_\mu\slash{\hspace{-5pt}\bar n}\gamma^\rho
\slash{\hspace{-5pt} n}\gamma^\sigma\slash{\hspace{-5pt}\bar n}\gamma_\nu/2$  and
$\hat N^{(2)\rho\sigma}_{\mu\nu}=q^-\gamma_\mu\gamma^\rho\slash{\hspace{-5pt}n}\gamma^\sigma\gamma_\nu$.
%$\hat{\bar N}^{(2)\rho\sigma}_{\mu\nu}=\gamma_\mu\gamma^\rho\slash{\hspace{-5pt}\bar n}\gamma^\sigma\gamma_\nu$.
We insert them into Eqs.(\ref{eq:Wsi0}-\ref{eq:Wsi2}) and obtain,
\begin{eqnarray}
\frac{d^2\tilde W^{(0)}_{\mu\nu}}{d^2k_\perp} &=&
\frac{1}{2}{\rm Tr}\bigl[\hat h^{(0)}_{\mu\nu}\hat\Phi^{(0)N}(x_B,k_\perp)\bigr],
\label{eq:W0} \\
%%%%%%%%%%%%%%%%%%%%%%%
\frac{d^2\tilde W^{(1,L)}_{\mu\nu}}{d^2k_\perp} &=&
\frac{1}{4q\cdot p}{\rm Tr}\bigl[\hat h^{(1)\rho}_{\mu\nu}\omega_\rho^{\ \rho'}\hat\varphi^{(1,L)N}_{\rho'}(x_B,k_\perp)\bigr],
\label{eq:W1L} \\
%%%%%%%%%%%%%%%%%
\frac{d^2\tilde W^{(2,L)}_{\mu\nu}}{d^2k_\perp} &=&
\frac{1}{(2q\cdot p)^2}\left\{{\rm Tr}\bigl[\hat h^{(1)\rho}_{\mu\nu}
\omega_\rho^{\ \rho'}\hat\phi^{(2,L)N}_{\rho'}(x_B,k_\perp)\bigr]
+
{\rm Tr}\bigl[\hat N^{(2)\rho\sigma}_{\mu\nu}\omega_\rho^{\ \rho'}\omega_\sigma^{\ \sigma'}
\hat\varphi^{(2,L)N}_{\rho'\sigma'}(x_B,k_\perp)\bigr]\right\},
\label{eq:W2L} \\
%%%%%%%%%%%%%%%%%%
\frac{d^2\tilde W^{(2,M)}_{\mu\nu}}{d^2k_\perp} &=&
\frac{1}{(2q\cdot p)^2}
{\rm Tr}\bigl[\hat h^{(2)\rho\sigma}_{\mu\nu}\omega_\rho^{\ \rho'}\omega_\sigma^{\ \sigma'}
\hat\varphi^{(2,M)N}_{\rho'\sigma'}(x_B,k_\perp)\bigr].
\label{eq:W2M}
\end{eqnarray}
The correlation matrices are defined as,
\begin{eqnarray}
\hat\varphi^{(1,L)N}_\rho(x_1,k_{1\perp})&\equiv&\int dx_2d^2k_{2\perp}\hat\Phi^{(1)N}_\rho(x_1,k_{1\perp},x_2,k_{2\perp}),\\
%%%%%%%%%%%%%
\hat\varphi^{(2,L)N}_{\rho\sigma}(x_1,k_{1\perp})&\equiv&\int dxd^2k_{\perp}
\frac{dx_2d^2k_{2\perp}}{x_2-x_1-i\varepsilon}
\hat\Phi^{(2)N}_{\rho\sigma}(x_1,k_{1\perp},x_2,k_{2\perp},x,k_\perp),\\
%%%%%%%%%%%%%%%%%%%%%
\hat\varphi^{(2,M)N}_{\rho\sigma}(x,k_{\perp})&\equiv&\int dx_1d^2k_{1\perp}dx_2d^2k_{2\perp}
\hat\Phi^{(2)N}_{\rho\sigma}(x_1,k_{1\perp},x_2,k_{2\perp},x,k_\perp),\\
%%%%%%%%%%%%%
\hat\phi^{(2,L)N}_\sigma(x_1,k_{1\perp})&\equiv&\int dxd^2k_\perp dx_2d^2k_{2\perp}
\hat\Phi^{(2)N}_{\rho\sigma}(x_1,k_{1\perp},x_2,k_{2\perp},x,k_\perp).
\end{eqnarray}
They are given by,
\begin{eqnarray}
&& \hat\varphi^{(1,L)N}_\rho(x,k_\perp)
=\int \frac{p^+dy^- d^2y_\perp}{(2\pi)^3}e^{ixp^+y^- -i\vec{k}_{\perp}\cdot \vec{y}_{\perp}}
\langle N|\bar\psi(0) D_\rho(0){\cal L}(0;y)\psi(y)|N\rangle,\\
%%%%%%%%%%%%%%%%
&&\hat\varphi^{(2,L)N}_{\rho\sigma}(x,k_{\perp})
=\int \frac{dx_2}{x_2-x-i\varepsilon}\frac{p^+dy^- d^2y_\perp}{(2\pi)^3}\frac{p^+dz^-}{2\pi}
e^{ix_2p^+z^-+ixp^+(y^--z^-)-i\vec{k}_{\perp}\cdot \vec{y}_\perp} \nonumber\\
&&\phantom{XXXXXXXXXXXXX}
\langle N|\bar\psi(0){\cal L}(0;z^-,y_\perp) D_\rho(z^-,y_\perp)D_\sigma(z^-,y_\perp){\cal L}(z^-,\vec y_\perp;y)\psi(y)|N\rangle,\\
%%%%%%%%%%%%%%%
&&\hat\varphi^{(2,M)N}_{\rho\sigma}(x,k_{\perp})
=\int \frac{p^+dy^- d^2y_\perp}{(2\pi)^3}e^{ixp^+y^- -i\vec{k}_\perp\cdot \vec{y}_{\perp}}
\langle N|\bar\psi(0) D_\rho(0){\cal L}(0;y)D_\sigma(y)\psi(y)|N\rangle,\\
%%%%%%%%%%%%%%%
&&\hat\phi^{(2,L)N}_\sigma(x,k_{\perp})
=\int \frac{p^+dy^- d^2y_\perp}{(2\pi)^3}e^{ixp^+y^- -i\vec{k}_\perp\cdot \vec{y}_{\perp}}
\langle N|\bar\psi(0) D^-(0)D_\sigma(0){\cal L}(0;y)\psi(y)|N\rangle.
\end{eqnarray}

We note that,
$\tilde W^{(0)*}_{\mu\nu}=\tilde W^{(0)}_{\nu\mu}$,
$\tilde W^{(2,M)*}_{\mu\nu}=\tilde W^{(2,M)}_{\nu\mu}$,
$\tilde W^{(1,R)}_{\mu\nu}=\tilde W^{(1,L)*}_{\nu\mu}$,
and $\tilde W^{(2,R)}_{\mu\nu}=\tilde W^{(2,L)*}_{\nu\mu}$.
Hence, if we divide $W_{\mu\nu}$ into a $\mu\leftrightarrow\nu$ symmetric part and an anti-symmetric part,
and denote $W_{\mu\nu}=W_{S,\mu\nu}+iW_{A,\mu\nu}$, we obtain,
\begin{equation}
\frac{d^2W_{S,\mu\nu}}{d^2k_\perp}=
\frac{d^2\tilde W^{(0)}_{S,\mu\nu}}{d^2k_\perp}+
2{\rm Re}\frac{d^2\tilde W^{(1,L)}_{S,\mu\nu}}{d^2k_\perp}+
2{\rm Re}\frac{d^2\tilde W^{(2,L)}_{S,\mu\nu}}{d^2k_\perp}+\frac{d^2\tilde W^{(2,M)}_{S,\mu\nu}}{d^2k_\perp},
\end{equation}
\begin{equation}
\frac{d^2W_{A,\mu\nu}}{d^2k_\perp}=
\frac{d^2\tilde W^{(0)}_{A,\mu\nu}}{d^2k_\perp}+
2{\rm Im}\frac{d^2\tilde W^{(1,L)}_{S,\mu\nu}}{d^2k_\perp}+
2{\rm Im}\frac{d^2\tilde W^{(2,L)}_{S,\mu\nu}}{d^2k_\perp}+\frac{d^2\tilde W^{(2,M)}_{A,\mu\nu}}{d^2k_\perp}.
\end{equation}

The  anti-symmetric part contributes only in reactions with polarized lepton.
In this paper, we concentrate on the unpolarized reactions and calculate the symmetric part in the following.

Now, we continue with a complete calculation of the hadronic tensor $d^2 W_{\mu \nu}/d^2k_\perp$
in the unpolarized $e^-+N\to e^-+q+X$ up to twist-4 level.
For this purpose, we need to calculate $d^2 W_{\mu \nu}/d^2k_\perp$ up to
$d^2\tilde W_{\mu \nu}^{(2)}/d^2k_\perp$ and we now present
the calculations of each term in the following.

The contribution from $d^2\tilde W_{\mu \nu}^{(0)}/d^2k_\perp$ is the easiest one to calculate.
Because $\hat H^{(0)}_{\mu\nu}(x)$ contains 3 $\gamma$-matrices, only $\gamma^\alpha$
term of $\hat\Phi^{(0)}(x,k_\perp)$ contributes in the unpolarized case so we need only to consider
$\hat\Phi^{(0)N}(x,k_\perp)=\gamma^\alpha\Phi^{(0)N}_\alpha(x,k_\perp)/2$,
\begin{equation}
\Phi^{(0)N}_{\alpha}(x,k_{\perp})
=\int \frac{p^+dy^- d^{2}y_{\perp}}{(2\pi)^3}
e^{ix p^+ y^- -i\vec{k}_{\perp}\cdot \vec{y}_{\perp}}
\langle N|\bar{\psi}(0)\frac{\gamma_\alpha}{2}{\cal{L}}(0;y)\psi(y)|N\rangle
=p_\alpha f_q^N+k_{\perp\alpha} f_{q\perp}^N+\frac{M^2}{p^+}n_\alpha f_{q(-)}^N.
\label{eq:Phi0alpha}
\end{equation}
and obtain the result for $d^2\tilde W_{\mu \nu}^{(0)}/d^2k_\perp$ as,
\begin{equation}
\frac{d^2\tilde W_{\mu \nu}^{(0)}}{d^2k_\perp}=-d_{\mu\nu}f_q^N(x_B,k_\perp)+
\frac{1}{q\cdot p}k_{\perp\{\mu}(q+x_Bp)_{\nu\}}f_{q\perp}^N(x_B,k_\perp)+
2(\frac{M}{q\cdot p})^2(q+x_Bp)_\mu (q+x_Bp)_\nu f_{q(-)}^N(x_B,k_\perp),
\end{equation}
where $d^{\mu\nu}=g^{\mu\nu}-\bar{n}^{\mu}n^{\nu}-\bar{n}^{\nu}n^{\mu}$ and
$A_{\{\mu}B_{\nu\}}\equiv A_\mu B_\nu+A_\nu B_\mu$,
$A_{[\mu}B_{\nu]}\equiv A_\mu B_\nu-A_\nu B_\mu$.
The TMD quark distribution/correlation functions are given by,
\begin{eqnarray}
&&f_q^N(x,k_\perp)=\frac{n^\alpha}{p^+}\Phi^{(0)N}_\alpha(x,k_\perp)=\int \frac{dy^- d^2y_{\perp}}{(2\pi)^3}
e^{ix p^+ y^- -i\vec{k}_{\perp}\cdot \vec{y}_{\perp}}
\langle N|\bar{\psi}(0)\frac{\gamma^+}{2}{\cal{L}}(0;y)\psi(y)|N\rangle,
\label{eq:fqn} \\
&& k_\perp^\alpha f_{q\perp}^N(x_B,k_\perp)=d^{\alpha\beta}\Phi^{(0)N}_\beta(x,k_\perp)
=\int \frac{p^+dy^- d^2y_{\perp}}{(2\pi)^3}
e^{ix p^+ y^- -i\vec{k}_{\perp}\cdot \vec{y}_{\perp}}
\langle N|\bar{\psi}(0)\frac{\gamma^\alpha_\perp}{2}{\cal{L}}(0;y)\psi(y)|N\rangle,
\label{eq:fqnperp} \\
&&f_{q(-)}^N(x_B,k_\perp)=\frac{p^+}{M^2}\bar n^\alpha\Phi^{(0)N}_\alpha(x,k_\perp)
=\frac{p^+}{M^2}\int \frac{p^+dy^- d^2y_{\perp}}{(2\pi)^3}
e^{ix p^+ y^- -i\vec{k}_{\perp}\cdot \vec{y}_{\perp}}
\langle N|\bar{\psi}(0)\frac{\gamma^-}{2}{\cal{L}}(0;y)\psi(y)|N\rangle.
\label{eq:fqnminus}
\end{eqnarray}

Because $\hat h^{(1)\rho}_{\mu\nu}$ contains 5 $\gamma$-matrices,
we have contributions from $\gamma_\alpha$ and $\gamma_5\gamma_\alpha$ terms of $\varphi^{(1,L)N}_\rho$,
i.e., we need to consider
%\begin{equation}
$\hat\varphi^{(1,L)N}_\rho(x,k_\perp)=[\gamma^\alpha\varphi^{(1)N}_{\rho\alpha}(x,k_\perp)-
\gamma_5\gamma^\alpha\tilde\varphi^{(1)N}_{\rho\alpha}(x,k_\perp)]/2
$  %\end{equation}
and obtain,
\begin{equation}
\frac{d^2\tilde W_{\mu\nu}^{(1,L)}}{d^2k_\perp}
=\frac{1}{2p\cdot q} \Bigl[h^{(1)\rho\alpha}_{\mu\nu}\omega_\rho^{\ \rho'}\varphi^{(1)N}_{\rho'\alpha}(x_B,k_\perp)
-\tilde h^{(1)\rho\alpha}_{\mu\nu}\omega_\rho^{\ \rho'}\tilde\varphi^{(1)N}_{\rho'\alpha}(x_B,k_\perp)\Bigr],
\end{equation}
where $h^{(1)\rho\alpha}_{\mu\nu}\equiv {\rm Tr}[\gamma^\alpha \hat h^{(1)\rho}_{\mu\nu}]/4$,
$\tilde h^{(1)\rho\alpha}_{\mu\nu}\equiv {\rm Tr}[\gamma_5\gamma^\alpha \hat h^{(1)\rho}_{\mu\nu}]/4$ and,
\begin{eqnarray}
&&\varphi^{(1)N}_{\rho\alpha}(x, k_\perp)
%={\rm Tr}\bigl[\frac{\gamma_\alpha}{2} \hat\varphi^{(1)N}_\rho(x,k_\perp)\bigr]
=\int \frac{p^+dy^-d^2y_\perp}{(2\pi)^3}e^{ixp^+y^--i\vec y_\perp\cdot\vec k_\perp}
\langle N|\bar\psi(0)\frac{\gamma_\alpha}{2}{\cal L}(0;y)D_\rho(y)\psi(y)|N\rangle,
\label{eq:varphi1}\\
&&\tilde\varphi^{(1)N}_{\rho\alpha}(x, k_\perp)
%={\rm Tr}\bigl[\frac{\gamma_5\gamma_\alpha}{2} \hat\varphi^{(1)N}_\rho(x,k_\perp)\bigr]
=\int \frac{p^+dy^-d^2y_\perp}{(2\pi)^3}e^{ixp^+y^--i\vec y_\perp\cdot\vec k_\perp}
\langle N|\bar\psi(0)\frac{\gamma_5\gamma_\alpha}{2}{\cal L}(0;y)D_\rho(y)\psi(y)|N\rangle.\
\label{eq:tildevarphi1}
\end{eqnarray}
After evaluating the two traces in $h^{(1)\rho\alpha}_{\mu\nu}$ and $\tilde h^{(1)\rho\alpha}_{\mu\nu}$, we
obtain the symmetric parts as,
\begin{eqnarray}
&&h^{(1)\rho\alpha}_{S,\mu\nu}=-g^\alpha_{\ \mu}d^\rho_{\ \nu}-g^\alpha_{\ \nu}d^\rho_{\ \mu}+g_{\mu\nu}d^{\rho\alpha},\\
&& \tilde h^{(1)\rho\alpha}_{S,\mu\nu}=ig^\alpha_{\ \mu}\varepsilon_{\perp\nu}^\rho
+ig^\alpha_{\ \nu}\varepsilon_{\perp\mu}^\rho-ig_{\mu\nu}\varepsilon_{\perp}^{\rho\alpha},
\end{eqnarray}
where $\epsilon_{\perp\rho\gamma}\equiv\epsilon_{\alpha\beta\rho\gamma}\bar n^\alpha n^\beta$.
%%%%%%
Up to twist-4, the contributing terms of $\varphi^{(1)N}_{\rho\alpha}(x,k_\perp)$
and  $\tilde\varphi^{(1)N}_{\rho\alpha}(x,k_\perp)$
are respectively,
\begin{eqnarray}
\varphi^{(1)N}_{\rho\alpha}(x,k_\perp)&=&
p_\alpha k_{\perp\rho}\varphi^{(1)N}_\perp(x,k_\perp)+
(k_{\perp\alpha}k_{\perp\rho}-\frac{k_\perp^2}{2}d_{\rho\alpha})\varphi^{(1)N}_{\perp 2}(x,k_\perp)+
%(n_\alpha k_{\perp\rho}+n_\rho k_{\perp\alpha})\varphi^{(1)N}_{\perp 3}(x,k_\perp)\nonumber\\
\frac{k_\perp^2}{2}(\bar n_{\{\alpha}n_{\rho\}}-d_{\rho\alpha})\varphi^{(1)N}_{\perp 3}(x,k_\perp),\\
\tilde\varphi^{(1)N}_{\rho\alpha}(x,k_\perp)&=&
ip_\alpha\varepsilon_{\perp\rho\gamma}k_\perp^\gamma \tilde\varphi^{(1)N}_\perp(x,k_\perp)+
\frac{i}{2}k_{\perp\{\alpha}\varepsilon_{\perp\rho\}\gamma}k_\perp^\gamma \tilde\varphi^{(1)N}_{\perp2}(x,k_\perp)+
\frac{i}{2}k_{\perp[\alpha}\varepsilon_{\perp\rho]\gamma}k_\perp^\gamma\tilde\varphi^{(1)N}_{\perp3}(x,k_\perp).
\end{eqnarray}
%where $k_{\perp\{\alpha}\varepsilon_{\perp\rho\}\gamma}$ and $k_{\perp[\alpha}\varepsilon_{\perp\rho]\gamma}$
%denote that we symmstrize or anti-symmetrize between $\rho$ and $\alpha$, i.e.,
%$k_{\perp\{\alpha}\varepsilon_{\perp\rho\}\gamma}\equiv
%k_{\perp\alpha}\varepsilon_{\perp\rho\gamma}+k_{\perp\rho}\varepsilon_{\perp\alpha\gamma}$ and
%$k_{\perp[\alpha}\varepsilon_{\perp\rho]\gamma}\equiv
%k_{\perp\alpha}\varepsilon_{\perp\rho\gamma}-k_{\perp\rho}\varepsilon_{\perp\alpha\gamma}$.
The result for $d^2\tilde W_{S,\mu\nu}^{(1,L)}/d^2k_\perp$ is,
\begin{eqnarray}
&&\frac{d^2\tilde W_{S,\mu\nu}^{(1,L)}}{d^2k_\perp}
=-\frac{1}{2q\cdot p}\bigl\{ (p_\mu k_{\perp\nu}+p_\nu k_{\perp\mu})
[\varphi^{(1)N}_{\perp}(x_B,k_\perp)-\tilde\varphi^{(1)N}_{\perp}(x_B,k_\perp)]\nonumber\\
&&\phantom{XXXXXXXX}
+(2k_{\perp\mu}k_{\perp\nu}-k_\perp^2d_{\mu\nu})
  [\varphi^{(1)N}_{\perp2}(x_B,k_\perp)-\tilde\varphi^{(1)N}_{\perp2}(x_B,k_\perp)]\nonumber\\
&&\phantom{XXXXXXXX}
%+(n_\mu k_{\perp\nu}+n_\nu k_{\perp\mu})\varphi^{(1)N}_{\perp3}(x_B,k_\perp)
+k_\perp^2(g_{\mu\nu}-d_{\mu\nu})
[\varphi^{(1)N}_{\perp3}(x_B,k_\perp)-\tilde\varphi^{(1)N}_{\perp3}(x_B,k_\perp)]\bigr\}.
\label{eq:W1Lres}
\end{eqnarray}

Up to twist-4 level, we need only to consider $\slash{\hspace{-5pt}p}$ and the $\gamma_5\slash{\hspace{-5pt}p}$-term
in the calculations of $d\tilde W^{(2)}_{\mu \nu}/d^2k_\perp$.
For the first term in Eq.(\ref{eq:W2L}), because of $\omega_\rho^{\ \rho'}$ and
$n_\rho\hat h^{(1)\rho}_{\mu\nu}=0$, we need only to consider the $k_{\perp\rho}$ terms and
we found out that they contribute only at twist-5 or higher level.
For the second term, because $n_\rho\hat N^{(2)\rho\sigma}_{\mu\nu}=n_\sigma\hat N^{(2)\rho\sigma}_{\mu\nu}=0$ and
$\hat\varphi^{(2,L)N}_{\rho\sigma}=\hat\varphi^{(2,L)N}_{\sigma\rho}$,
we need to consider only $k_{\perp\rho}k_{\perp\sigma}$ and $k_\perp^2d_{\rho\alpha}$
for the tensor term and $k_{\perp\{\rho}\varepsilon_{\perp\sigma\}\gamma}k_\perp^\gamma$
for the pseudo-tensor term.
Furthermore,
\begin{eqnarray}
&&k^2_\perp\hat N^{(2)\rho\sigma}_{\mu\nu}d_{\rho\sigma}=2\hat N^{(2)\rho\sigma}_{\mu\nu}k_{\perp\rho}k_{\perp\sigma}
=-2k_\perp^2\gamma_\mu\slash{\hspace{-5pt}n}\gamma_\nu,\\
&&\hat N^{(2)\rho\sigma}_{\mu\nu}k_{\perp\rho}\varepsilon_{\perp\sigma\gamma}k_\perp^\gamma=
-\hat N^{(2)\rho\sigma}_{\mu\nu}k_{\perp\sigma}\varepsilon_{\perp\rho\gamma}k_\perp^\gamma
=k_\perp^2\gamma_\mu\slash{\hspace{-5pt}n}\slash{\hspace{-5pt}n_{\perp1}}\slash{\hspace{-5pt}n_{\perp2}}\gamma_\nu,
\end{eqnarray}
we need only to consider,
\begin{equation}
\hat\varphi^{(2,L)N}_{\rho\sigma}(x,k_\perp)=\frac{\slash{\hspace{-5pt}p}}{2}
\left(-\frac{1}{2}k^2_\perp d_{\rho\sigma}\right)\varphi^{(2,L)N}_\perp(x,k_\perp)+...
\end{equation}
and obtain the results for $d^2\tilde W^{(2,L)}_{\mu\nu}/d^2k_\perp$ up to $1/Q^2$ as,
\begin{equation}
\frac{d^2\tilde W^{(2,L)}_{\mu\nu}}{d^2k_\perp}=
-\frac{1}{2q\cdot p}k_\perp^2d_{\mu\nu}\varphi^{(2,L)N}_\perp(x_B,k_\perp)+... .
\end{equation}

Similarly, to calculate $d^2\tilde W^{(2,M)}_{\mu\nu}/d^2k_\perp$ up to $1/Q^2$ level, we need to consider
\begin{equation}
\hat\varphi^{(2,M)N}_{\rho\sigma}(x,k_\perp)=
\frac{\slash{\hspace{-5pt}p}}{2}\left(-\frac{1}{2}k^2_\perp d_{\rho\sigma}\right)\varphi^{(2,M)N}_\perp(x,k_\perp)-
\frac{i}{4}\gamma_5\slash{\hspace{-5pt}p}
k_{\perp[\rho}\varepsilon_{\perp\sigma]\gamma}k_\perp^\gamma\tilde\varphi^{(2,M)N}_\perp(x,k_\perp),
\end{equation}
and the results for  $d^2\tilde W^{(2,M)}_{\mu\nu}/d^2k_\perp$ are given by,
\begin{equation}
\frac{d^2\tilde W^{(2,M)}_{\mu\nu}}{d^2k_\perp}=
\frac{k_\perp^2}{(q\cdot p)^2}p_\mu p_\nu[\varphi^{(2,M)N}_\perp(x_B,k_\perp)-\tilde\varphi^{(2,M)N}_\perp(x_B,k_\perp)].
\end{equation}

%%%%%%%%%%%%%%%
QCD equation of motion relates matrix elements with different number of $D_\rho$ and gives
\begin{eqnarray}
&&xf_{q\perp}^N(x,k_\perp)=-[\varphi_\perp^{(1)N}(x,k_\perp)-\tilde\varphi_\perp^{(1)N}(x,k_\perp)],\\
%&&x\varphi_{\perp3}^N(x,k_\perp)=-\phi_\perp^{(2,L)N}(x,k_\perp)\textcolor{blue}{-}\tilde\phi_\perp^{(2,L)N}(x,k_\perp),\\
&&2(xM)^2f_{q(-)}^N(x,k_\perp)=k_\perp^2[\varphi_\perp^{(2,M)N}(x,k_\perp)-\tilde\varphi_\perp^{(2,M)N}(x,k_\perp)],\\
&&x[\varphi^{(1)N}_{\perp 3}(x,k_\perp)-\tilde \varphi^{(1)N}_{\perp 3}(x,k_\perp)]
=-[\varphi^{(2,M)N}_{\perp }(x,k_\perp)-\tilde \varphi^{(2,M)N}_{\perp }(x,k_\perp)],
\end{eqnarray}
where, as well as in the following of this paper,
all the correlation functions in the results of the hadronic tensors and/or cross section
stand for their real parts. %
The final results for $d^2W_{\mu\nu}/d^2k_\perp$ up to twist-4 level are given by,
 \begin{eqnarray}
\frac{d^2W_{\mu \nu}}{d^2k_\perp}&=&
-\frac{1}{q\cdot p}\Bigl\{ (q\cdot p)d_{\mu\nu}f_q^N(x_B,k_\perp)
+\frac{2M^2}{q\cdot p}(q+2x_Bp)_\mu(q+2x_Bp)_\nu f_{q(-)}^N(x_B,k_\perp)\nonumber\\
&&-(q+2x_Bp)_{\{\mu} k_{\perp\nu\}} f_{q\perp}^N(x_B,k_\perp) +(2k_{\perp\mu}k_{\perp\nu}-k_\perp^2d_{\mu\nu})
  [\varphi^{(1)N}_{\perp2}(x_B,k_\perp)-\tilde\varphi^{(1)N}_{\perp2}(x_B,k_\perp)]\nonumber\\
&&+k_\perp^2d_{\mu\nu} \varphi^{(2,L)N}_{\perp2}(x_B,k_\perp)\Bigr\}.
\end{eqnarray}

\section{Differential cross section and $\langle \cos 2\phi\rangle$ up to the $1/Q^2$}

Making the Lorentz contraction of the result for $d^2W_{\mu\nu}/d^2k_\perp$
with the leptonic tensor $L_{\mu\nu}$ given in Eq.(3),
we obtain the differential cross section as,
\begin{eqnarray}
\frac{d\sigma}{dx_Bdyd^2k_\perp}&=&\frac{2\pi\alpha_{em}^2e_q^2}{Q^2y}
\bigg\{ [1+(1-y)^2] f_q^N(x_B,k_\perp)
-4(2-y)\sqrt{1-y} \frac{|\vec k_\perp|}{Q} x_Bf_{q\perp}^{(1)N}(x_B,k_\perp)\cos\phi \nonumber\\
\nonumber &&-4(1-y)\frac{|\vec k_\perp|^2}{Q^2}x_B
[\varphi^{(1)N}_{\perp 2}(x_B,k_\perp)-\tilde\varphi^{(1)N}_{\perp 2} (x_B,k_\perp)]\cos 2\phi\\
\nonumber &&+8(1-y)\left(\frac{|\vec k_\perp|^2}{Q^2}
x_B[\varphi^{(1)N}_{\perp 2}(x_B,k_\perp)-\tilde\varphi^{(1)N}_{\perp 2} (x_B,k_\perp)]
+\frac{2x_B^2M^2}{Q^2}f_{q(-)}^N (x_B,k_\perp)\right)\\
&&-2\left[1+(1-y)^2\right]\frac{|\vec k_\perp|^2}{Q^2}x_B\varphi_{\perp 2}^{(2,L)N} (x_B,k_\perp)\bigg\}.
\label{eq:CSres}
\end{eqnarray}
%where the azimuthal angle $\phi$ is defined by $\cos\phi=\hat{\vec{k}}_\perp\cdot \hat{\vec{l}}_\perp$,

\end{widetext}

From Eq.(\ref{eq:CSres}), we can calculate the azimuthal asymmetries
$\langle\cos\phi\rangle$ and $\langle\cos 2\phi\rangle$.
The result for $\langle\cos\phi\rangle$ and its nuclear dependence are
discussed in \cite{Gao:2010mj}.
We now discuss the result for $\langle\cos 2\phi\rangle$.
At fixed $k_\perp$, it is given by,
\begin{eqnarray}
&&\langle \cos2\phi\rangle_{eN}=-\frac{2(1-y)}{1+(1-y)^2}\frac{|\vec k_\perp|^2}{Q^2}\times\phantom{XXXXXXXX}\nonumber\\
&&\phantom{XXXX}\frac{x_B[\varphi_{\perp 2}^{(1)N}(x_B,k_\perp)-\tilde\varphi_{\perp 2}^{(1)N}(x_B,k_\perp)]}{f_q^N(x_B,k_\perp)}.\ \
\label{eq:cos2phires}
\end{eqnarray}
Integrating over the magnitude of $\vec k_\perp$, we obtain,
\begin{eqnarray}
&&\langle\langle \cos 2\phi\rangle\rangle_{eN}=
-\frac{2(1-y)}{1+(1-y)^2} \times\phantom{XXXX}\nonumber\\
&&\phantom{XX}\frac{\int |\vec k_\perp|^2d^2k_\perp
x_B[\varphi_{\perp 2}^{(1)N}(x_B,k_\perp)-\tilde\varphi_{\perp 2}^{(1)N}(x_B,k_\perp)]}{Q^2f_q^N(x_B)},\nonumber
\end{eqnarray}
where $f_q^N(x)=\int d^2k_\perp f_q^N(x,k_\perp)$ is the usual quark distribution in nucleon.
The new quark correlation functions involved are given by,
\begin{eqnarray}
|\vec k_\perp|^2\varphi_{\perp 2}^{(1)N}(x,k_\perp)&=(2\hat k_\perp^\alpha\hat k_\perp^\rho+d^{\alpha\rho})
\varphi^{(1)N}_{\rho\alpha}(x,k_\perp), \\
|\vec k_\perp|^2\tilde\varphi_{\perp 2}^{(1)N}(x,k_\perp)&=
-i\hat k_\perp^{\{\alpha}\varepsilon_\perp^{\rho\}\sigma}\hat k_{\perp\sigma} \tilde\varphi^{(1)N}_{\rho\alpha}(x,k_\perp),
\end{eqnarray}
where $\hat k_\perp=k_\perp/|\vec k_\perp|$ denotes the unit vector.
If we consider only ``free parton with intrinsic transverse momentum",
i.e., the same case as considered in \cite{Cahn:1978se},
we need to just set $g=0$ in the results mentioned above.
In this case, ${\cal L}=1$ and
$x[\varphi_{\perp 2}^{(1)N}(x,k_\perp)-\tilde\varphi_{\perp 2}^{(1)N}(x,k_\perp)]=f_q^N(x,k_\perp)$,
so that,
\begin{equation}
\langle \cos 2\phi\rangle_{eN}|_{g=0}=-\frac{2(1-y)}{1+(1-y)^2}\frac{|\vec k_\perp|^2}{Q^2},
\label{eq:cos2phig=0}
\end{equation}
which is just the result obtained in \cite{Cahn:1978se}.

In general, we need to take QCD multiple parton scattering into account thus
$\langle \cos 2\phi\rangle_{eN}$ is given by Eq.~(\ref{eq:cos2phires}) where new
quark correlation functions are involved.
Measurements of $\langle \cos 2\phi\rangle_{eN}$, in particular whether the
results deviate from  Eq.~(\ref{eq:cos2phig=0}),  can provide useful information
on the new parton correlation functions and on multiple parton scattering as well.

%\section{$A$-dependence of the azimuthal asymmetry}

If we consider $e^-+A\to e^-+q+X$, i.e. instead of a nucleon but a nucleus target,
all the calculations given above apply and we obtain similar results
with only a replacement of  the state $|N\rangle$
by $|A\rangle$ in the definitions of the matrix elements and/or parton distribution/correlation functions.
The multiple gluon scattering now can be connected to different nucleons in the nucleus $A$
thus give rise to nuclear dependence.
It has been shown that,  under the ``maximal two gluon approximation",
a TMD quark distribution $\Phi^A_\alpha(x,k_\perp)$ in nucleus defined in the form,
\begin{eqnarray}
\Phi^A_\alpha(x,k_\perp)&\equiv &\int \frac{p^+dy^-d^2y_\perp}{(2\pi)^3}
e^{ixp^+y^- -i\vec  k_\perp\cdot \vec y_\perp}\times\nonumber\\
&&\langle A \mid \bar\psi(0)\Gamma_\alpha{\cal L}(0;y)\Psi(y)\mid A \rangle,
\label{form}
\end{eqnarray}
is given by a convolution of the corresponding distribution $\Phi^N_\alpha(x,k_\perp)$ in nucleon
and a Gaussian broadening,
\begin{equation}
\Phi^A_\alpha(x,k_\perp)\approx\frac{A}{\pi \Delta_{2F}}
\int d^2\ell_\perp e^{-(\vec k_\perp -\vec\ell_\perp)^2/\Delta_{2F}}\Phi^N_\alpha(x,\ell_\perp),
\label{tmdgeneral}
\end{equation}
where $\Gamma_\alpha$ is any gamma matrix,  $\Psi(y)$ is a field operator;
$\Delta_{2F}$  is the broadening width given by,
\begin{equation}
\Delta_{2F}=\int d\xi^-_N \hat q_F(\xi_N)=\frac{2\pi^2\alpha_s}{N_c}\int d\xi^-_N\rho_N^A(\xi_N)[xf^N_g(x)]_{x=0},
\label{eq:Delta2F}
\end{equation}
%$\hat q_F(\xi_N)$ is the quark transport parameter,
%\begin{equation}
%\hat q_F(\xi_N)
%=-\frac{g^2}{2N_c}\rho_N^A(\xi_N)\int \frac{d\xi^-}{2p^+}\langle N \mid F_{+\sigma}(0)F_+^{\sigma}(\xi^-)\mid N \rangle
%=\frac{2\pi^2\alpha_s}{N_c}\rho_N^A(\xi_N)[xf^N_g(x)]_{x=0},
%\label{qhat1}
%\end{equation}
where $\rho_N^A(\xi_N)$ is the spatial nucleon number density inside the nucleus
and $f^N_g(x)$ is the gluon distribution function in nucleon.
% which is defined as,
%\begin{equation}
%xf^N_g(x)=\int\frac{d\xi^-}{2\pi p^+} e^{ixp^+\xi^-}
%\langle N\mid F^{+\sigma}(0){F_{\sigma}}^+(\xi^-)\mid N\rangle.
%\end{equation}

We note that both $\varphi_{\rho\alpha}(x,k_\perp)$ and  $\tilde\varphi_{\rho\alpha}(x,k_\perp)$
have the form of $\Phi^A_\alpha(x,k_\perp)$.
Hence,
\begin{equation}
\varphi^{(1)A}_{\rho\alpha}(x,k_\perp)
\approx\frac{A}{\pi \Delta_{2F}}
\int d^2\ell_\perp e^{-(\vec k_\perp -\vec\ell_\perp)^2/\Delta_{2F}}\varphi_{\rho\alpha}^{(1)N}(x,\ell_\perp),
\label{tmdphi}
\end{equation}
\begin{equation}
\tilde\varphi^{(1)A}_{\rho\alpha}(x,k_\perp)
\approx\frac{A}{\pi \Delta_{2F}}
\int d^2\ell_\perp e^{-(\vec k_\perp -\vec\ell_\perp)^2/\Delta_{2F}}\tilde\varphi_{\rho\alpha}^{(1)N}(x,\ell_\perp),
\label{tmdtildephi}
\end{equation}
Making the Lorentz contration of both sides of these two equations %Eqs. (\ref{tmdphi}) and (\ref{tmdtildephi})
with $2\hat k_\perp^\rho \hat k_\perp^\alpha+d^{\rho\alpha}$ and $\hat k_\perp^{\{\alpha}\varepsilon_\perp^{\rho\}\sigma}$
respectively, we obtain that,
\begin{eqnarray}
&&|\vec k_\perp|^2\varphi^{(1)A}_{\perp2}(x,k_\perp)
\approx \frac{A}{\pi \Delta_{2F}}
\int d^2\ell_\perp e^{-(\vec k_\perp -\vec\ell_\perp)^2/\Delta_{2F}}\times\nonumber\\
&&\phantom{XXXXX}
\left[2(\ell_\perp\cdot\hat k_\perp)^2+\ell_\perp^2\right]\varphi_{\perp2}^{(1)N}(x,\ell_\perp),\\
\label{eq:phi1NA}
&&|\vec k_\perp|^2\tilde\varphi^{(1)A}_{\perp2}(x,k_\perp)
\approx \frac{A}{\pi \Delta_{2F}}
\int d^2\ell_\perp e^{-(\vec k_\perp -\vec\ell_\perp)^2/\Delta_{2F}}\times\nonumber\\
&&\phantom{XXXXX}
\left[2(\ell_\perp\cdot\hat k_\perp)^2+\ell_\perp^2\right]\tilde\varphi_{\perp2}^{(1)N}(x,\ell_\perp).
\label{eq:tildephi1NA}
\end{eqnarray}
Adopting a Gaussian ansatz, i.e.,
\begin{eqnarray}
f_q^N(x,k_\perp)&=&\frac{1}{\pi\alpha}f_q^N(x)e^{-\vec k_\perp^2/\alpha},\\
\varphi_{\perp2}^{(1)N}(x,k_\perp)&=&\frac{1}{\pi\beta}\varphi_{\perp2}^{(1)N}(x)e^{-\vec k_\perp^2/\beta},\\
\tilde\varphi_{\perp2}^{(1)N}(x,k_\perp)&=&\frac{1}{\pi\tilde\beta}\tilde\varphi_{\perp2}^{(1)N}(x)e^{-\vec k_\perp^2/\tilde\beta},
\end{eqnarray}
we obtain, for those functions in nucleus,
\begin{equation}
f_q^A(x,k_\perp)\approx\frac{A}{\pi\alpha_A}f_q^N(x)e^{-\vec k_\perp^2/\alpha_A},
\end{equation}
\begin{equation}
\varphi_{\perp2}^{(1)A}(x,k_\perp)\approx
\frac{A}{\pi\beta_A}\Bigl(\frac{\beta}{\beta_A}\Bigr)^2
\varphi_{\perp2}^{(1)N}(x)e^{-\vec k_\perp^2/\beta_A},
\end{equation}
\begin{equation}
\tilde\varphi_{\perp2}^{(1)A}(x,k_\perp)\approx
\frac{A}{\pi\tilde\beta_A}\Bigl(\frac{\tilde\beta}{\tilde\beta_A}\Bigr)^2
\tilde\varphi_{\perp2}^{(1)N}(x)e^{-\vec k_\perp^2/\tilde\beta_A},
\end{equation}
where $\alpha_A=\alpha+\Delta_{2F}$, $\beta_A=\beta+\Delta_{2F}$ and $\tilde\beta_A=\tilde\beta+\Delta_{2F}$.
The azimuthal asymmetry is given by,
%\begin{equation}
%\langle \cos 2\phi\rangle_{eN}=-\frac{2(1-y)}{1+(1-y)^2}\frac{x_B\vec k_\perp^2}{Q^2}
%\frac{\frac{1}{\beta}\varphi_{\perp2}^{(1)N}(x_B)e^{-\vec k_\perp^2/\beta}
%-\frac{1}{\tilde\beta}\tilde\varphi_{\perp2}^{(1)N}(x_B)e^{-\vec k_\perp^2/\tilde\beta}}
%{\frac{1}{\alpha}f_q^N(x_B)e^{-\vec k_\perp^2/\alpha}},
%\label{eq:cos2phieN}
%\end{equation}
%
%\begin{equation}
%\langle \cos 2\phi\rangle_{eA}=-\frac{2(1-y)}{1+(1-y)^2}\frac{x_B\vec k_\perp^2}{Q^2}
%\frac{
%\frac{\beta^2}{\beta_A^3}\varphi_{\perp2}^{(1)N}(x_B)e^{-\vec k_\perp^2/\beta_A}
%-\frac{\tilde\beta^2}{\tilde\beta_A^3}
%\tilde\varphi_{\perp2}^{(1)N}(x_B)e^{-\vec k_\perp^2/\tilde\beta_A} }
%{\frac{1}{\alpha_A}f_q^N(x_B)e^{-\vec k_\perp^2/\alpha_A}},
%\label{eq:cos2phieA}
%\end{equation}
%
\begin{eqnarray}
&&\frac{\langle \cos 2\phi\rangle_{eA}}{\langle \cos 2\phi\rangle_{eN}}\approx
\frac{\alpha}{\alpha_A}e^{-{\vec k_\perp}^2/\alpha_A+\vec k_\perp^2/\alpha}\times \nonumber\\
&&\phantom{XX}\frac{
\frac{\beta^2}{\beta_A^3}\varphi_{\perp2}^{(1)N}(x_B)e^{-\vec k_\perp^2/\beta_A}
-\frac{\tilde\beta^2}{\tilde\beta_A^3}
\tilde\varphi_{\perp2}^{(1)N}(x_B)e^{-\vec k_\perp^2/\tilde\beta_A}  }
{\Bigl[\frac{1}{\beta}\varphi_{\perp2}^{(1)N}(x_B)e^{-\vec k_\perp^2/\beta}
-\frac{1}{\tilde\beta}\tilde\varphi_{\perp2}^{(1)N}(x_B)e^{-\vec k_\perp^2/\tilde\beta}\Bigr]}, \nonumber
\end{eqnarray}
which reduces to
%\begin{equation}
%\langle \cos 2\phi\rangle_{eN}=-\frac{2(1-y)}{1+(1-y)^2}\frac{x_B\vec k_\perp^2}{Q^2}
%\frac{\varphi_{\perp2}^{(1)N}(x_B)-\tilde\varphi_{\perp2}^{(1)N}(x_B)}{f_q^N(x_B)},
%\label{eq:cos2phieN2}
%\end{equation}
%
%\begin{equation}
%\langle \cos2\phi\rangle_{eA}=-\frac{2(1-y)}{1+(1-y)^2}\frac{x_B\vec k_\perp^2}{Q^2}\left(\frac{\beta}{\beta_A}\right)^2
%\frac{ \varphi_{\perp2}^{(1)N}(x_B)-\tilde\varphi_{\perp2}^{(1)N}(x_B)}
%{f_q^N(x_B)},
%\label{eq:cos2phieA2}
%\end{equation}
\begin{equation}
\frac{\langle \cos 2\phi\rangle_{eA}}{\langle \cos 2\phi\rangle_{eN}}
=(\frac{\beta}{\beta+\Delta_{2F}})^2,
\end{equation}
in the case that $\alpha=\beta=\tilde\beta$.
We see that, in this case, for given $x_B$, $Q^2$ and $|\vec k_\perp|$,
$\langle\cos2\phi\rangle_{eA}$ in deep inelastic $eA$ scattering
is suppressed compared to that in $eN$ scattering with a
suppression factor $\beta^2/(\beta+\Delta_{2F})^2$.
Comparing with result of ~\cite{Gao:2010mj}, we can see that $\langle\cos2\phi\rangle_{eA}$ is more suppressed
than $\langle\cos\phi\rangle_{eA}$.
In general, $\beta,\tilde \beta$ can be different from $\alpha$, and the ratio can also be different
at different $k_\perp$ and $\Delta_{2F}$.
As example, we show the results for a few cases in Figs. 1a and 1b with $\beta=\tilde \beta$.
 \begin{figure} [htb]%1
 \resizebox{0.45\textwidth}{!}{\includegraphics{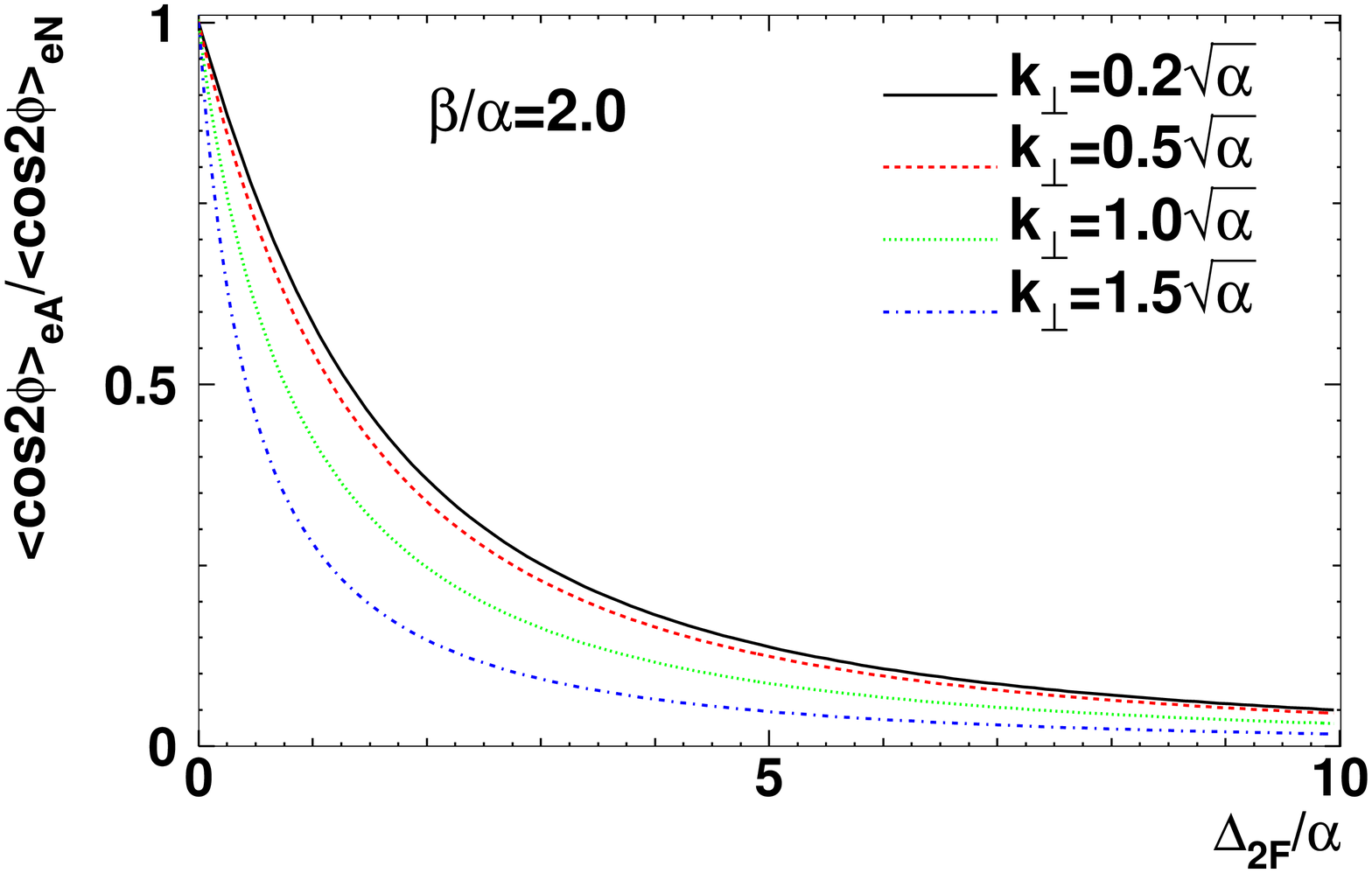}}
 \resizebox{0.45\textwidth}{!}{\includegraphics{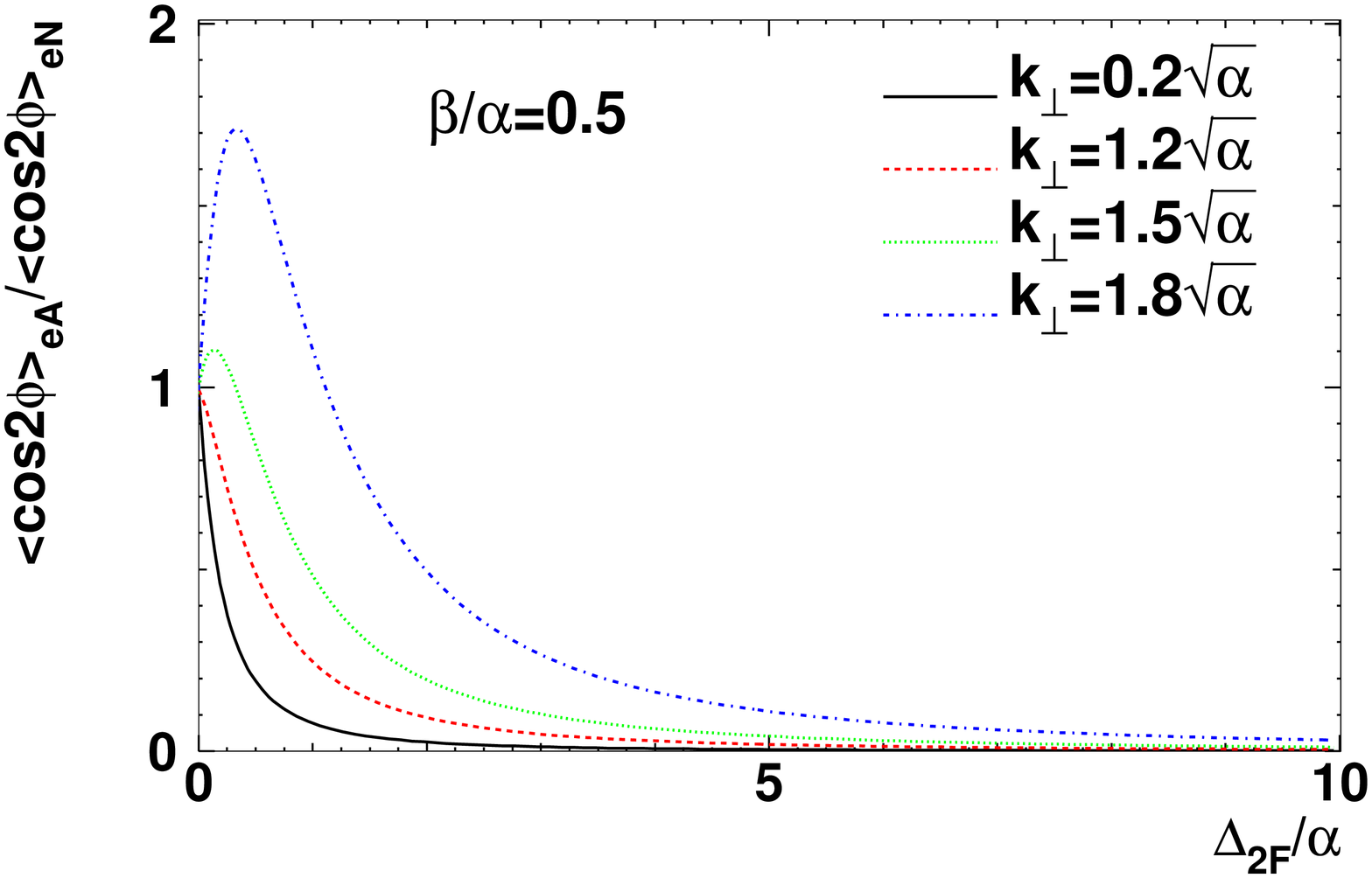}}
    \caption{(color online) Ratio $\langle \cos2\phi\rangle_{eA}/\langle \cos2\phi\rangle_{eN}$ as a function of $\Delta_{2F}$ for different $k_\perp$ and $\beta$.  }
    \label{fig:AfH}       % Give a unique label
    \end{figure}

We see that  the asymmetry can be suppressed or enhanced depending on the values of  $k_\perp$ and $\Delta_{2F}$,
and the magnitude is smaller than $\langle\cos\phi\rangle$case.

If we integrate over the magnitude of $\vec k_\perp$, we obtain,
%\begin{equation}
%\int d^2k_\perp \vec k_\perp^2\varphi_{\perp 2}^{(1)N}(x,k_\perp)=\beta\varphi_{\perp 2}^{(1)N}(x),
%\end{equation}
%\begin{equation}
%\int d^2k_\perp \vec k_\perp^2\tilde\varphi_{\perp 2}^{(1)N}(x,k_\perp)= \tilde\beta\tilde\varphi_{\perp2}^{(1)N}(x),
%\end{equation}
%\begin{equation}
%\int d^2k_\perp \vec k_\perp^2\varphi_{\perp 2}^{(1)A}(x,k_\perp)\approx
%A(\frac{\beta}{\beta_A})^2\beta_A\varphi_{\perp2}^{(1)N}(x),
%\end{equation}
%\begin{equation}
%\int d^2k_\perp \vec k_\perp^2\tilde\varphi_{\perp 2}^{(1)A}(x,k_\perp)\approx
%A(\frac{\tilde\beta}{\tilde\beta_A})^2\tilde\beta_A\tilde\varphi_{\perp2}^{(1)N}(x),
%\end{equation}
%\begin{equation}
%\langle\langle \cos 2\phi\rangle\rangle_{eN}=-\frac{2(1-y)}{1+(1-y)^2}\frac{x_B}{Q^2}
%\frac{\beta\varphi_{\perp2}^{(1)N}(x_B)-\tilde\beta\tilde\varphi_{\perp2}^{(1)N}(x_B)}{f_q^N(x_B)},
%\end{equation}
%\begin{equation}
%\langle\langle \cos 2\phi\rangle\rangle_{eA}\approx-\frac{2(1-y)}{1+(1-y)^2}\frac{x_B}{Q^2}
%\frac{(\frac{\beta}{\beta_A})^2\beta_A\varphi_{\perp2}^{(1)N}(x_B)
%-(\frac{\tilde\beta}{\tilde\beta_A})^2\tilde\beta_A\tilde\varphi_{\perp2}^{(1)N}(x_B)}{f_q^N(x_B)},
%\end{equation}
%
\begin{equation}
\frac{\langle\langle \cos\phi\rangle\rangle_{eA}}{\langle\langle \cos\phi\rangle\rangle_{eN}}
\approx\frac{(\frac{\beta}{\beta_A})^2\beta_A\varphi_{\perp2}^{(1)N}(x_B)
-(\frac{\tilde\beta}{\tilde\beta_A})^2\tilde\beta_A\tilde\varphi_{\perp2}^{(1)N}(x_B)}
{\beta\varphi_{\perp2}^{(1)N}(x_B)-\tilde\beta\tilde\varphi_{\perp2}^{(1)N}(x_B)},
\end{equation}
which reduces to $\beta/(\beta+\Delta_{2F})$ for the special case $\beta=\tilde\beta$.

%\end{widetext}

\section{Summary and discussions}

We calculated the hadronic tensor and differential cross section for unpolarized SIDIS process
$e^-+N\to e^-+q+X$ in LO pQCD and up to twist-4 contributions
The results depend on a number of new TMD parton correlation functions.
We showed that measurements of the azimuthal asymmetry $\langle \cos2\phi\rangle$
and its $k_\perp$-dependence provides information on these TMD
correlation  functions which in turn can shed light on the properties of multiple gluon interaction
in hadronic processes. Under two-gluon correlation approximation, we also show the relationship between these TMD
correlation functions inside large nuclei and that of a nucleon. One can therefore study the
nuclear dependence of the azimuthal asymmetry $\langle \cos2\phi\rangle$ which
is determined by the jet transport parameter $\hat q$ inside nuclei. With 
a Gaussian ansatz for the TMD parton correlation functions inside the nucleon, we also illustrate 
numerically that the asymmetry $\langle \cos2\phi\rangle$ is suppressed 
in the corresponding SIDIS with nuclear target.

There exist experimental measurements of the azimuthal asymmetries in both unpolarized and polarized 
DIS \cite{Aubert:1983cz,Arneodo:1986cf,Adams:1993hs,Breitweg:2000qh,Chekanov:2002sz,
Airapetian:2001eg,Airapetian:2002mf,Airapetian:2004tw,Alexakhin:2005iw,Webb:2005cd,:2008rv,Alekseev:2010dm}.
More results are expected from CLAS at JLab and COMPASS at CERN.
The available data seem to be consistent with the Gaussian ansatz for the 
transverse momentum dependence of the TMD matrix elements\cite{Schweitzer:2010tt}.
However these data are still not adequate enough to provide any precise  constraints on the form of 
the higher twist matrix elements.  Our calculations of the azimuthal asymmetries are
most valid in the small transverse momentum region where NLO pQCD corrections are
not dominant. The high twist effects are also most accessible in intermediate region
of $Q^{2}$.  One expects that future experiments such as those at the proposed Electron Ion 
Collider (EIC) \cite{EIC} will be better equipped to study these high twist effects in detail.

This work was supported in part by the National Natural Science Foundation of China
under the approval Nos. 10975092 and 11035003,
the China Postdoctoral Science Foundation funded project under
Contract No.20090460736,
Office of Energy
Research, Office of High Energy and Nuclear Physics, Division of
Nuclear Physics, of the U.S. Department of Energy under Contract No.
DE-AC02-05CH11231.


\begin{thebibliography}{}

%\cite{Georgi:1977tv}
\bibitem{Georgi:1977tv}
  H.~Georgi and H.~Politzer,
  %``Clean Tests Of QCD In Mu P Scattering,''
  Phys.\ Rev.\ Lett.\  {\bf 40}, 3 (1978).
  %%CITATION = PRLTA,40,3;%%

%\cite{Cahn:1978se}
\bibitem{Cahn:1978se}
  R.~N.~Cahn,
  %``Azimuthal Dependence In Leptoproduction:  A Simple Parton Model
  %Calculation,''
  Phys.\ Lett.\ B {\bf 78}, 269 (1978).
  %%CITATION = PHLTA,B78,269;%%

%\cite{Berger:1979kz}
\bibitem{Berger:1979kz}
 E.~L.~Berger,
  %``Higher Twist Effects In Deep Inelastic Scattering,''
 Phys.\ Lett.\ B {\bf 89}, 241 (1980).
  %%CITATION = PHLTA,B89,241;%%

%\cite{Liang:1993re}
\bibitem{Liang:1993re}
%  Z.~Liang and B.~Nolte-Pautz,
 Z.~T.~Liang and B.~Nolte-Pautz,
%``Azimuthal asymmetry in polarized leptoproduction and spin dependent
%structure functions,''
 Z.\ Phys.\ C {\bf 57}, 527 (1993).
%%CITATION = ZEPYA,C57,527;%%

%\cite{Mulders:1995dh}
\bibitem{Mulders:1995dh}
  P.~J.~Mulders and R.~D.~Tangerman,
  %``The complete tree-level result up to order 1/Q for polarized deep-inelastic leptoproduction,''
  Nucl.\ Phys.\ B {\bf 461}, 197 (1996)
  [Erratum {\bf 484}, 538 (1997)].

%\cite{Oganesian:1997jq}
\bibitem{Oganesian:1997jq}
 K.~A.~Oganesian, H.~R.~Avakian, N.~Bianchi and P.~Di Nezza,
  %``Investigations of azimuthal asymmetry in semi-inclusive  leptoproduction,''
  Eur.\ Phys.\ J.\ C {\bf 5}, 681 (1998).
%  [arXiv:hep-ph/9709342].
  %%CITATION = HEP-PH 9709342;%%

%\cite{Chay:1997qy}
\bibitem{Chay:1997qy}
  J.~Chay and S.~M.~Kim,
  %``Azimuthal correlation in lepton hadron scattering via charged  weak-current processes,''
  Phys.\ Rev.\ D {\bf 57}, 224 (1998)
  [arXiv:hep-ph/9705284].
  %%CITATION = HEP-PH 9705284;%%

%\cite{Nadolsky:2000kz}
\bibitem{Nadolsky:2000kz}
  P.~M.~Nadolsky, D.~R.~Stump and C.~P.~Yuan,
  %``Azimuthal asymmetries at HERA: Theoretical aspects,''
  Phys.\ Lett.\  B {\bf 515}, 175 (2001)
  [arXiv:hep-ph/0012262].
  %%CITATION = PHLTA,B515,175;%%


%\cite{Anselmino:2006rv}
\bibitem{Anselmino:2006rv}
  M.~Anselmino, M.~Boglione, A.~Prokudin and C.~Turk,
  %``Semi-inclusive deep inelastic scattering processes from small to large
  %P(T),''
  Eur.\ Phys.\ J.\  A {\bf 31}, 373 (2007)
  [arXiv:hep-ph/0606286].
  %%CITATION = EPHJA,A31,373;%%

%\cite{Anselmino:2005nn}
\bibitem{Anselmino:2005nn}
  M.~Anselmino, M.~Boglione, U.~D'Alesio, A.~Kotzinian, F.~Murgia and A.~Prokudin,
  %``The role of Cahn and Sivers effects in deep inelastic scattering,''
  Phys.\ Rev.\  D {\bf 71}, 074006 (2005)
  [arXiv:hep-ph/0501196].
  %%CITATION = PHRVA,D71,074006;%%
  
%\cite{Liang:2006wp}
\bibitem{Liang:2006wp}
  Z.~T.~Liang and X.~N.~Wang,
  %``Azimuthal and single spin asymmetry in deep-inelastic lepton-nucleon
  %scattering,''
  Phys.\ Rev.\  D {\bf 75}, 094002 (2007)
  [arXiv:hep-ph/0609225].
  %%CITATION = PHRVA,D75,094002;%%

%\cite{Gao:2010mj}
\bibitem{Gao:2010mj}
  J.~H.~Gao, Z.~T.~Liang and X.~N.~Wang,
  %``Nuclear dependence of azimuthal asymmetry in semi-inclusive deep inelastic
  %scattering,''
  Phys.\ Rev.\  C {\bf 81}, 065211 (2010)
  [arXiv:1001.3146 [hep-ph]].
  %%CITATION = PHRVA,C81,065211;%%

%\cite{Aubert:1983cz}
\bibitem{Aubert:1983cz}
  J.~J.~Aubert {\it et al.}  [European Muon Collaboration],
  %``Measurement Of Hadronic Azimuthal Distributions In Deep Inelastic Muon
  %Proton Scattering,''
  Phys.\ Lett.\ B {\bf 130}, 118 (1983).
  %%CITATION = PHLTA,B130,118;%%
%
%\cite{Arneodo:1986cf}
\bibitem{Arneodo:1986cf}
  M.~Arneodo {\it et al.}  [European Muon Collaboration],
  %``Measurement Of Hadron Azimuthal Distributions In Deep Inelastic Muon Proton
  %Scattering,''
  Z.\ Phys.\ C {\bf 34}, 277 (1987).
  %%CITATION = ZEPYA,C34,277;%%

%\cite{Adams:1993hs}
\bibitem{Adams:1993hs}
  M.~R.~Adams {\it et al.}  [E665 Collaboration],
  %``Perturbative QCD effects observed in 490-GeV deep inelastic muon
  %scattering,''
  Phys.\ Rev.\ D {\bf 48}, 5057 (1993).
  %%CITATION = PHRVA,D48,5057;%%

%\cite{Breitweg:2000qh}
\bibitem{Breitweg:2000qh}
  J.~Breitweg {\it et al.}  [ZEUS Collaboration],
  %``Measurement of azimuthal asymmetries in deep inelastic scattering,''
  Phys.\ Lett.\ B {\bf 481}, 199 (2000).
%  [arXiv:hep-ex/0003017].
  %%CITATION = HEP-EX 0003017;%%
%
%\cite{Chekanov:2002sz}
\bibitem{Chekanov:2002sz}
  S.~Chekanov {\it et al.}  [ZEUS Collaboration],
  %``Study of the azimuthal asymmetry of jets in neutral current deep inelastic
  %scattering at HERA,''
  Phys.\ Lett.\ B {\bf 551}, 226 (2003).
%  [arXiv:hep-ex/0210064].
  %%CITATION = HEP-EX 0210064;%%
%
%\cite{Airapetian:2001eg}
\bibitem{Airapetian:2001eg}
  A.~Airapetian {\it et al.}  [HERMES Collaboration],
  %``Single-spin azimuthal asymmetries in electroproduction of neutral pions  in
  %semi-inclusive deep-inelastic scattering,''
  Phys.\ Rev.\  D {\bf 64}, 097101 (2001)
  [arXiv:hep-ex/0104005].
  %%CITATION = PHRVA,D64,097101;%%

%

%\cite{Airapetian:2002mf}
\bibitem{Airapetian:2002mf}
  A.~Airapetian {\it et al.}  [HERMES Collaboration],
  %``Measurement of single-spin azimuthal asymmetries in semi-inclusive
  %electroproduction of pions and kaons on a longitudinally polarised deuterium
  %target,''
  Phys.\ Lett.\  B {\bf 562}, 182 (2003)
  [arXiv:hep-ex/0212039].
  %%CITATION = PHLTA,B562,182;%%

%\cite{Airapetian:2004tw}
\bibitem{Airapetian:2004tw}
 A.~Airapetian {\it et al.}  [HERMES Collaboration],
%   ``Single-spin asymmetries in semi-inclusive deep-inelastic scattering on  a
%   transversely polarized hydrogen target,''
 Phys.\ Rev.\ Lett.\  {\bf 94}, 012002 (2005).
%  [arXiv:hep-ex/0408013].
 %%CITATION = HEP-EX 0408013;%%

%\cite{Alexakhin:2005iw}
\bibitem{Alexakhin:2005iw}
V.~Y.~Alexakhin {\it et al.}  [COMPASS Collaboration],
%   ``First measurement of the transverse spin asymmetries of the deuteron in
%   semi-inclusive deep inelastic scattering,''
 Phys.\ Rev.\ Lett.\  {\bf 94}, 202002 (2005).
%  [arXiv:hep-ex/0503002].
 %%CITATION = HEP-EX 0503002;%%
%
%\cite{Webb:2005cd}
\bibitem{Webb:2005cd}
R.~Webb  [COMPASS Collaboration],
%   ``First measurements of Collins and Sivers asymmetries at COMPASS,''
 Nucl.\ Phys.\ A {\bf 755}, 329 (2005).
%  [arXiv:hep-ex/0501031].
%%CITATION = HEP-EX 0501031;%%

%\cite{:2008rv}
\bibitem{:2008rv}
  M.~Osipenko {\it et al.}  [CLAS Collaboration],
  %``Measurement of unpolarized semi-inclusive pi+ electroproduction off the
  %proton,''
  Phys.\ Rev.\  D {\bf 80}, 032004 (2009)
  [arXiv:0809.1153 [hep-ex]].
  %%CITATION = PHRVA,D80,032004;%%


%\cite{Alekseev:2010dm}
\bibitem{Alekseev:2010dm}
  M.~G.~Alekseev {\it et al.},
  %``Azimuthal asymmetries of charged hadrons produced by high-energy muons
  %scattered off longitudinally polarised deuterons,''
  arXiv:1007.1562 [hep-ex].
  %%CITATION = ARXIV:1007.1562;%%

%\cite{Boer:1997nt}
\bibitem{Boer:1997nt}
  D.~Boer and P.~J.~Mulders,
  %``Time-reversal odd distribution functions in leptoproduction,''
  Phys.\ Rev.\  D {\bf 57}, 5780 (1998)
  [arXiv:hep-ph/9711485].
  %%CITATION = PHRVA,D57,5780;%%


%\cite{Liang:2008vz}
\bibitem{Liang:2008vz}
  Z.~T.~Liang, X.~N.~Wang and J.~Zhou,
  %``The Transverse-momentum-dependent Parton Distribution Function and Jet
  %Transport in Medium,''
  Phys.\ Rev.\  D {\bf 77}, 125010 (2008).
  %[arXiv:0801.0434 [hep-ph]].
  %%CITATION = PHRVA,D77,125010;%%
    
  
  %\cite{VanHaarlem:2007kj}
\bibitem{VanHaarlem:2007kj}
  Y.~Van Haarlem, A.~Jgoun and P.~Di Nezza,
%``Nuclear p(t) broadening at HERMES,''
{\it In the Proceedings of 9th Workshop on Non-Perturbative Quantum Chromodynamics, Paris, France, 4-8 Jun 2007, pp 10}
  [arXiv:0704.3712 [hep-ex]].
 %%CITATION = ECONF,C0706044,10;%%
%\cite{VanHaarlem:2009zz}
\bibitem{VanHaarlem:2009zz}
  Y.~Van Haarlem,
 %``Nuclear medium effects on hadronization measured by the Hermes collaboration ,''
  Nucl.\ Phys.\ Proc.\ Suppl.\  {\bf 186}, 106 (2009).
 %%CITATION = NUPHZ,186,106;%%

%\cite{Domdey:2008aq}
\bibitem{Domdey:2008aq}
 S.~Domdey, D.~Grunewald, B.Z.~Kopeliovich and H.J.~Pirner,
%``Transverse Momentum Broadening in Semi-inclusive DIS on Nuclei,''
 Nucl.\ Phys.\  A {\bf 825}, 200 (2009) [arXiv:0812.2838 [hep-ph]].
 %%CITATION = NUPHA,A825,200;%%

  
 %\cite{Gyulassy:1993hr}
\bibitem{Gyulassy:1993hr}
  M.~Gyulassy and X.~N.~Wang,
  %``Multiple collisions and induced gluon Bremsstrahlung in QCD,''
  Nucl.\ Phys.\  B {\bf 420}, 583 (1994).
  %%[arXiv:nucl-th/9306003].
  %%CITATION = NUPHA,B420,583;%%

%\cite{Baier:1996sk}
\bibitem{Baier:1996sk}
  R.~Baier, Y.~L.~Dokshitzer, A.~H.~Mueller, S.~Peigne and D.~Schiff,
   %``Radiative energy loss and p(T)-broadening of high energy partons in nuclei,''
  Nucl.\ Phys.\  B {\bf 484}, 265 (1997).
  %%[arXiv:hep-ph/9608322].
  %%CITATION = NUPHA,B484,265;%%

%\cite{Wiedemann:2000za}
\bibitem{Wiedemann:2000za}
  U.~A.~Wiedemann,
   %``Gluon radiation off hard quarks in a nuclear environment: Opacity expansion,''
  Nucl.\ Phys.\  B {\bf 588}, 303 (2000)
  %%[arXiv:hep-ph/0005129].
  %%CITATION = NUPHA,B588,303;%%

%\cite{Gyulassy:2000er}
\bibitem{Gyulassy:2000er}
  M.~Gyulassy, P.~Levai and I.~Vitev,
  %``Reaction operator approach to non-Abelian energy loss,''
  Nucl.\ Phys.\  B {\bf 594}, 371 (2001).
  %%[arXiv:nucl-th/0006010].
  %%CITATION = NUPHA,B594,371;%%

%\cite{Guo:2000nz}
\bibitem{Guo:2000nz}
  X.~F.~Guo and X.~N.~Wang,
   %``Multiple Scattering, Parton Energy Loss and Modified Fragmentation
  %Functions in Deeply Inelastic eA Scattering,''
  Phys.\ Rev.\ Lett.\  {\bf 85}, 3591 (2000).
  %%[arXiv:hep-ph/0005044].
  %%CITATION = PRLTA,85,3591;%%

%\cite{Wang:2001ifa}
\bibitem{Wang:2001ifa}
  X.~N.~Wang and X.~F.~Guo,
  %``Multiple parton scattering in nuclei: Parton energy loss,''
  Nucl.\ Phys.\  A {\bf 696}, 788 (2001).
  %%[arXiv:hep-ph/0102230].
  %%CITATION = NUPHA,A696,788;%%
  
  
  %\cite{McLerran:1993ka}
\bibitem{McLerran:1993ka}
  L.~D.~McLerran and R.~Venugopalan,
  %``Gluon distribution functions for very large nuclei at small transverse
  %momentum,''
  Phys.\ Rev.\  D {\bf 49}, 3352 (1994)
  [arXiv:hep-ph/9311205].
  %%CITATION = PHRVA,D49,3352;%%



%\cite{Ellis:1982wd}
\bibitem{Ellis:1982wd}
  R.~K.~Ellis, W.~Furmanski and R.~Petronzio,
  %``Power Corrections To The Parton Model In QCD,''
  Nucl.\ Phys.\  B {\bf 207}, 1 (1982); and
  %%CITATION = NUPHA,B207,1;%%
%
%\cite{Ellis:1982cd}
%\bibitem{Ellis:1982cd}
 % R.~K.~Ellis, W.~Furmanski and R.~Petronzio,
  %``Unraveling Higher Twists,''
  Nucl.\ Phys.\  B {\bf 212}, 29 (1983).
  %%CITATION = NUPHA,B212,29;%%

%\cite{Qiu:1988dn}
\bibitem{Qiu:1988dn}
 J.~W.~Qiu,
%``TWIST FOUR CONTRIBUTIONS TO THE PARTON STRUCTURE FUNCTIONS,''
 Phys.\ Rev.\  D {\bf 42}, 30 (1990).
%%CITATION = PHRVA,D42,30;%%

%\cite{Qiu:1990xxa}
\bibitem{Qiu:1990xxa}
  J.~W.~Qiu and G.~F.~Sterman,
%   ``Power corrections in hadronic scattering. 1. Leading 1/Q**2 corrections to the Drell-Yan cross-section,''
  Nucl.\ Phys.\  B {\bf 353}, 105 (1991); and
  %%CITATION = NUPHA,B353,105;%%
%
%\cite{Qiu:1990xy}
%\bibitem{Qiu:1990xy}
%  J.~w.~Qiu and G.~F.~Sterman,
%``Power corrections to hadronic scattering. 2. Factorization,''
  Nucl.\ Phys.\  B {\bf 353}, 137 (1991).
%%CITATION = NUPHA,B353,137;%%


%\cite{Anselmino:2008sga}
\bibitem{Anselmino:2008sga}
  M.~Anselmino {\it et al.},
  %``Sivers Effect for Pion and Kaon Production in Semi-Inclusive Deep Inelastic
  %Scattering,''
  Eur.\ Phys.\ J.\  A {\bf 39}, 89 (2009)
  [arXiv:0805.2677 [hep-ph]].
  %%CITATION = EPHJA,A39,89;%%

%\cite{Schweitzer:2010tt}
\bibitem{Schweitzer:2010tt}
  P.~Schweitzer, T.~Teckentrup and A.~Metz,
  %``Intrinsic transverse parton momenta in deeply inelastic reactions,''
  Phys.\ Rev.\  D {\bf 81}, 094019 (2010)
  [arXiv:1003.2190 [hep-ph]].
  %%CITATION = PHRVA,D81,094019;%%


\bibitem{EIC} See e.g., 
EIC Collaboration, A High Luminosity, High Energy Electron-Ion-Collider  - A White Paper Prepared 
for the NSAC LRP 2007, 24 April, 2007; 
%\cite{Deshpande:2005wd}
%\bibitem{Deshpande:2005wd}
  A.~Deshpande, R.~Milner, R.~Venugopalan and W.~Vogelsang,
  %``Study of the fundamental structure of matter with an electron ion
  %collider,''
  Ann.\ Rev.\ Nucl.\ Part.\ Sci.\  {\bf 55}, 165 (2005)
  [arXiv:hep-ph/0506148].
  %%CITATION = ARNUA,55,165;%%


 \end{thebibliography}
\end{document}